\documentclass[desactivate]{aa}  

\usepackage{graphicx}
\usepackage[varg]{txfonts}
\usepackage{booktabs}
\usepackage{chemformula}
\let\ce\ch

\usepackage{orcidlink}
\usepackage{pifont}
\usepackage{cleveref}

\begin{document} 

    \titlerunning{Resolved Gas Temperatures and \ce{^{12}C/^{13}C} ratios in SVS13A via \ce{CH3CN} and \ce{CH3^{13}CN}.}
   \title{Resolved Gas Temperatures and \ce{^{12}C/^{13}C} ratios in SVS13A from ALMA Observations of \ce{CH3CN} and \ce{CH3^{13}CN}.}
    

   \author
   {T.-H. Hsieh
          \inst{1,2,3}\orcidlink{0000-0002-5507-5697}
          \and          
          J. E. Pineda
          \inst{1}\orcidlink{0000-0002-3972-1978}
          \and
            D. M. Segura-Cox
          \inst{4,1}\orcidlink{0000-0003-3172-6763}
          \and
          P. Caselli
          \inst{1}\orcidlink{0000-0003-1481-7911}
          \and
          M. J. Maureira
          \inst{1}\orcidlink{0000-0002-7026-8163}
          \and
          L. A. Busch
          \inst{1}
          \and
          M. T. Valdivia-Mena
          \inst{1,5}\orcidlink{0000-0002-0347-3837}
          \and
          C. Gieser
          \inst{1}\orcidlink{0000-0002-8120-1765}
          \and
          Y. Lin
          \inst{1}
          \and
          Y.-R. Chou
          \inst{1}
          \and
          V. Lattanzi
          \inst{1}\orcidlink{0000-0001-9819-1658}
          \and
          S. Spezzano
          \inst{1}\orcidlink{0000-0002-6787-5245}
          \and
          A. Lopez-Sepulcre
          \inst{6,7}
          \and
          R. Neri
          \inst{6}\orcidlink{0000-0002-7176-4046}
          }

   \institute{Max-Planck-Institut f\"{u}r extraterrestrische Physik, Giessenbachstrasse 1, D-85748 Garching, Germany \\ \email{thhsieh@asiaa.sinica.edu.tw}
    \and
    Taiwan Astronomical Research Alliance (TARA), Taiwan
    \and
    Institute of Astronomy and Astrophysics, Academia Sinica, P.O. Box 23-141, Taipei 106, Taiwan 
         \and
            Department of Physics and Astronomy, University of Rochester, Rochester, NY 14627, USA
        \and 
            European Southern Observatory, Karl-Schwarzschild-Strasse 2 85748 Garching bei Munchen, Munchen, Germany
        \and
            Institut de Radioastronomie Millim\'{e}trique (IRAM), 300 rue de la Piscine, F-38406, Saint-Martin d'H\`{e}res, France
        \and
            Univ. Grenoble Alpes, CNRS, IPAG, 38000 Grenoble, France
             }

   \date{Feb 1, 2023}

 
  \abstract
   {Multiple systems are common in field stars, and the frequency is found to be higher in early evolutionary stages. 
   Thus, the study of young multiple systems during the embedded stages is key to have a comprehensive understanding of star formation. In particular, how material accretes from the large-scale envelope into the inner region and how this flow interacts with the system physically and chemically has not yet been well characterized observationally.}
   {We aim to provide a snapshot of the forming protobinary system SVS13A, consisting of VLA4A and VLA4B. This includes clear pictures of its kinematic structures, physical conditions, and chemical properties.}
   {We conducted ALMA observations toward SVS13A targeting \ce{CH3CN} and \ce{CH3^{13}CN} J=12-11 K-ladder line emission with a high spatial resolution of $\sim30$ au at a spectral resolution of $\sim0.08$ km s$^{-1}$}
   {We perform LTE radiative transfer models to fit the spectral features of the line emission. We find the two-layer LTE radiative model including dust absorption is essential to interpret the \ce{CH3CN} and \ce{CH3^{13}CN} line emission. We identify two major and four small kinematic components, and derive their physical and chemical properties.}
   {We find a possible infalling signature toward the bursting secondary source VLA4A, which may be fed by an infalling streamer from the large-scale envelope. The mechanical heating in the binary system, as well as the infalling shocked gas, likely play a role in the thermal structure of the protobinary system. By accumulating mass from the streamer, the system might have experienced a gravitationally unstable phase before the accretion outburst. Finally, the derived \ce{CH3CN}/\ce{CH3^{13}CN} ratio is lower than the canonical ratio in the ISM and is different between VLA4A and VLA4B.}

   \keywords{ISM:kinematics and dynamics --
                ISM: individual objects: SVS13A --
                stars: protostars--
                stars: formation
               }

   \maketitle
%
\section{Introduction} \label{sec:introduction}
Observations reveal that the frequency of multiple stellar systems increases during earlier evolutionary stages, with a multiplicity rate of 0.57 for Class 0 and 0.23 for Class I \citep{to18,of23}. Multiple systems are also commonly observed among field stars \citep{to14}.
Thus, to understand the most common pathway of star formation, it is essential to study multiple systems at an early stage (e.g., \citealt{mu24}).

In contrast to the classical picture that stars form through isotropic collapse of dense cores \citep{sh77,te98}, recent molecular line observations reveal non-axisymmetric accretion in progress \citep{pi23}.
This infalling material is distributed in spatially elongated yet narrow structures, named ``infalling streamers'', that have been seen at different evolutionary stages \citep{ye19,pi20,al20,gi21,mu22b,ca21,ga22,th22,va22,fl23,gu24,va24}.
These infalling streamers can singificantly contribute to the mass accretion rate from the core to disk, $\sim$10$^{-6}$ M$_\odot$ yr$^{-1}$ \citep{va22}. Infalling streamers are also seen in numerical simulation such as \citet{ku17}.
In addition, infalling streamers might play a crucial role in changing the physical structures and chemical compositions in the inner region \citep{ga22,po24}.
The complex organic molecule (COM) \ce{CH3CN}, with its K-ladder transitions, is widely used as a thermometer toward hot dense regions. It is more commonly seen in high-mass star-forming regions but is also found in low-mass star-forming regions \citep{bo07,bi08,be18, lo18,ca18,be20,ya21}.

SVS13A is a well studied protobinary system located in the NGC1333 cluster in the Perseus Molecular Cloud ($d=$293 pc, \citealt{or18}). 
It is a Class I close binary consisting of VLA4A and VLA4B (or Per-emb-44B and 44A) with a separation of $0\farcs3$ ($\sim$90 au) based on continuum emission \citep{an04,to18,se18}. 
\citet{di22} derive the total mass of the binary to be 1.0$\pm$0.4 $M_\odot$ given the proper motion of the binary orbit. The individual masses of the protostars are estimated as 0.27$\pm$0.10 $M_\odot$ for VLA4A (secondary) and 0.60$\pm$0.20 $M_\odot$ for VLA4B (primary). The SVS13A binary system has a high bolometric luminosity of $L_{\rm bol}=45.3~L_\odot$
likely because the secondary VLA4A is in an outburst phase \citep{hs19}. In addition, \citet{hs23} found a large-scale ($\sim$700 au) candidate streamer connected with the dusty spiral linked to VLA4A. This streamer is possibly infalling, funneling material to the binary system with a mass infall rate $>1.4\times10^{-6} M_\odot$ yr$^{-1}$.
\ce{CH3CN} and the isotopologues \ce{CH2DCN} and \ce{CH3^{13}CN} have been previously detected toward SVS13A \citep{bi19,hs23,bi22a}. However, the emission had not been spatially resolved. 
Observations of O-bearing COMs at $\sim$300 au resolution in \citet{hs24} indicated that multiple structures should exist at tens of au scales based on multiple velocity components.

In this work, we present new Atacama Large Millimeter
ter/submillimeter Array (ALMA) molecular line observations of \ce{CH3CN} and \ce{CH3^{13}CN} towards SVS13A. We resolve the emission from the COM species \ce{CH3CN} and \ce{CH3^{13}CN} down to 30 au (0.1 arcsecond).
The high angular- and spectral-resolution data resolved the line emission, tracing the disks or rotating envelope around the protobinary VLA4A and VLA4B.
In Section \ref{sec:observation}, we present the data and imaging process. The results are reported in Section \ref{sec:result}. In Section \ref{sec:analysis}, we discuss the LTE radiative transfer fitting and identify the kinematic components. The discussions and conclusions are in Sections \ref{sec:discussion} and \ref{sec:summary}, respectively.

\begin{figure*}
\includegraphics[width=0.99\textwidth]{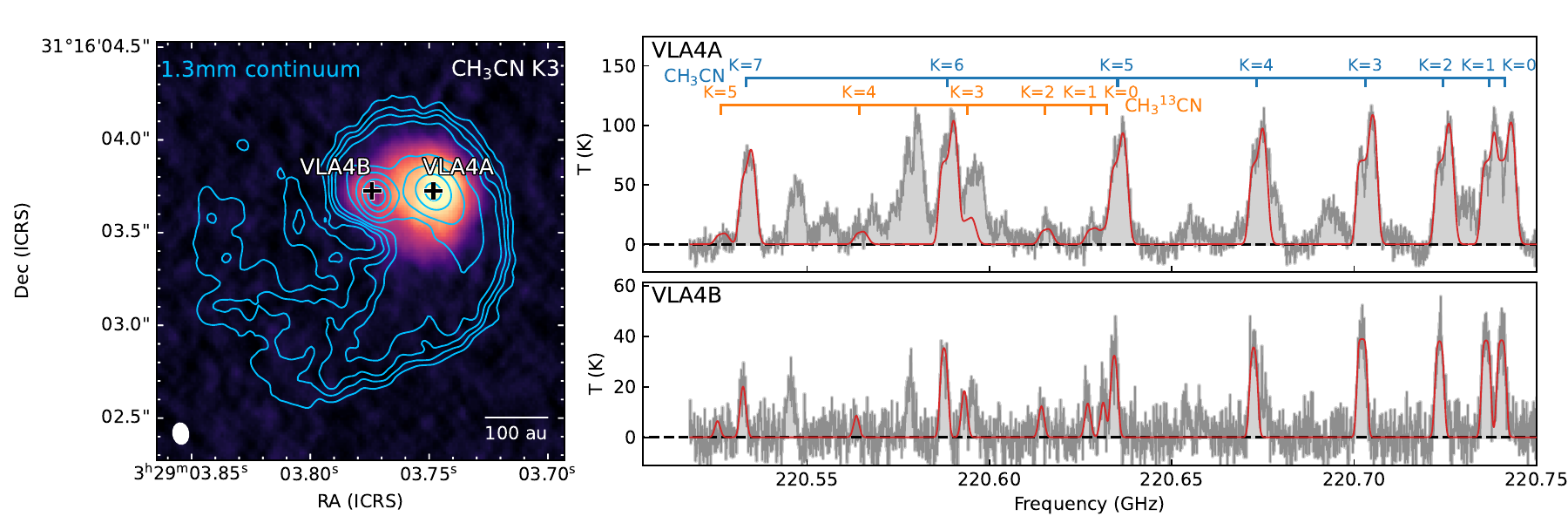}
\caption{(Left) Integrated intensity map of the \ce{CH3CN} J=12-11 K=3 ladder with the contours from 1.3 mm continuum at levels of 3, 5, 7, 10, 20, 40, 100, 200$\sigma$ with $\sigma=0.2$ K. (Right) The spectra are from the cross-marked positions of VLA4A (top) and VLA4B (bottom) from the \ce{CH3CN} window. The red lines represent the best-fit from our models (see sections \ref{sec:one-layer} and \ref{sec:two-layers}).}
\label{fig:spe_vla}
\end{figure*}

\section{Observations} \label{sec:observation}
\subsection{ALMA observation}
We conducted ALMA observations in June 2023 with three executions with the project ID 2022.1.00479.
In these executions, 43, 44, 45 antennas were used. PWV among the observations were from 0.4 to 0.55 mm. With the ALMA C-7 configuration, the baselines were $91-8282$ m ($\sim$$65-6000$ k$\lambda$) at 220 GHz, recovering scales from 0.1 to 1.2 arcsec ($\sim$30$-$360 au). The total on-source (SVS13A) time was 144 min toward the phase center at International Celestial Reference System (ICRS) $\alpha$ = 3$^{\rm h}$29$^{\rm m}$ 03$^{\rm s}$.74, $\delta$ = 31$^{\rm d}$16$^{\rm m}$03$^{\rm s}$.8. The three executions shared the same bandpass and flux calibrator as J0423-0120, while the phase calibrator either J0328+3139 or J0336-3218 depending on the execution.

The observations were designed to observe multiple chemical and kinematic tracers toward SVS13A such that multiple high-resolution spectral windows were placed accompanied with one continuum window at 234.5 GHz with 1.875 GHz bandwidth and 488 kHz (0.6 km s$^{-1}$) channel width (band 6, \citealt{ed04}). In this paper, we present the \ce{CH3CN} J=12-11 window which covers a wavelength ranging from 220.52 GHz to 220.75 GHz with a channel width of 61 kHz (0.08 km s$^{-1}$).
We use the pipeline calibrated data restored using CASA 6.4.12 \citep{casa}. Imaging was done using the {\tt tclean} package with a robust weighting of 0.5, resulting in a beam size of $0\farcs11\times0\farcs08$ ($\sim$30 au) for the \ce{CH3CN} cube. This image cube has an rms noise level of 1.7 mJy beam$^{-1}$ (4.7 K) at the spectral resolution of 0.08 km s$^{-1}$.

\subsection{STATCONT for continuum subtraction}
Many spectral lines adding up to a 'forest' features have been seen in high-mass star forming regions \citep{sa18}, making the determination of the continuum level difficult. This may also happen in low-mass star forming regions for the chemically-rich sources, e.g., hot corinos \citep{ce04}, especially with high-sensitivity data. SVS13A, with high luminosity (45.3 L$_\odot$), has plenty of line features already known in the literature \citep{bi19,bi22,di22,hs23}, see also Figure \ref{fig:spe_vla}.

We use STATCONT \citep{sa18} to determine the continuum level and find the line-free channels. STATCONT performs continuum subtraction in image domain pixel by pixel, and provides a continuum-subtracted cube and continuum image as outputs.
We used the default sigma-clipping algorithm, which removes the line-emitting channels by iteratively measuring the median and standard deviation in a histogram of spectral points. As a result, we obtained the continuum image from the continuum window and the line image cube from the \ce{CH3CN} window.

\section{Results} \label{sec:result}
\subsection{Continuum emission}
Figure \ref{fig:spe_vla} shows the continuum image as contours. The binary components VLA4A as the secondary and VLA4B as the primary are well resolved and two spirals are seen, consistent with previous observations \citep{to18,di22}. 
The continuum emission from the secondary (VLA4A) is more extended than that of the primary. However, VLA4A peak brightness is lower (70 K) than the primary VLA4B (90K).
The relative brightness is consistent with that at 0.9 mm, 260 K for VLA4A and 430 K for VLA4B \citep{di22}.
It is likely that the inner 30 au are optically thick for both VLA4A and VLA4B at 0.9 mm, as seen in other compact binary system (e.g., \citealt{ma22}). In this case, the difference could indicate either a lower dust temperature for the secondary or higher amounts of material therein, leading to higher levels of self-obscuration (e.g., \citealt{ga18}).
VLA4A likely is the more active source during an accretion outburst \citep{hs19}, also given its connection to the potential infalling streamer (the spiral) and the spectral features to be discussed below.

\subsection{\ce{CH3CN} and \ce{CH3^{13}CN}}
\ce{CH3CN} and \ce{CH3^{13}CN} J $=12-11$ emission are spatially resolved toward SVS13A. The K-ladder is detected in the observing windows with K $=0-7$ for \ce{CH3CN} and K $=0-5$ for \ce{CH3^{13}CN} (Figure \ref{fig:spe_vla}). 
Given the forest of lines, a particular line can be severely contaminated by other molecules especially for \ce{CH3^{13}CN}. \ce{CH3CN} emission is relatively strong but is still partially affected by a few bright lines such as \ce{CH3OCHO} and \ce{C2H5OH} (see Figure 2 in \citealt{hs23}). Using the broad wavelength coverage provided by the large program PROtostars \& DIsks: Global Evolution (PRODIGE) using NOEMA, \citet{hs23, hs24} thoroughly identify the molecular lines by performing LTE modeling for each molecule with multiple transitions. However, there are still unidentified lines in the spectral window presented here (Figure \ref{fig:spe_vla}). 

\begin{figure*}
\begin{center}
\includegraphics[width=0.97\textwidth]{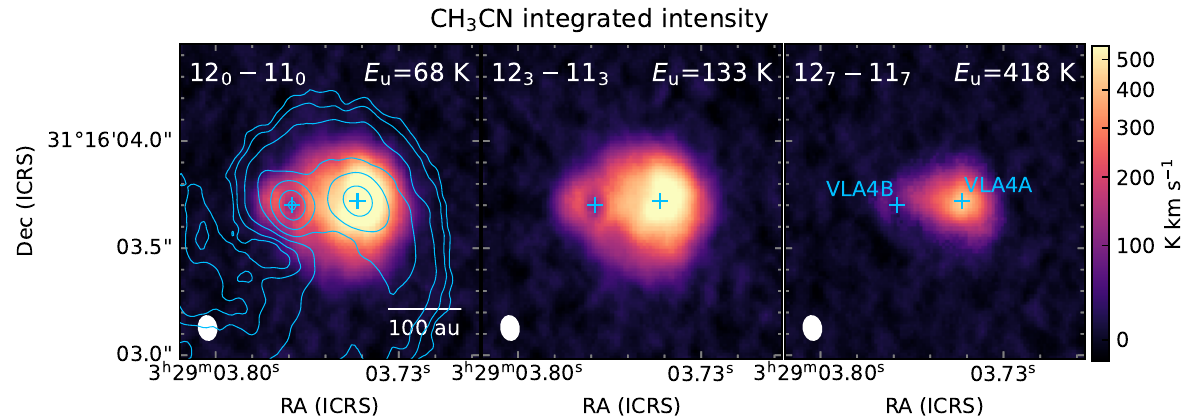}
\end{center}
\caption{Integrated intensity map of the \ce{CH3CN} J=12-11 with K=0, 3, and 7. The integration range is 4-12 km s$^{-1}$. The contours represent the 1.3 mm continuum emission at levels of 3, 5, 10, 20, 50, 150, and 300$\sigma$ with $\sigma$=0.2 K.
The blue cross markers indicate the locations of VLA4A and VLA4B from the continuum emission.
}
\label{fig:12_mom0}
\end{figure*}

\begin{figure*}
\begin{center}
\includegraphics[width=0.97\textwidth]{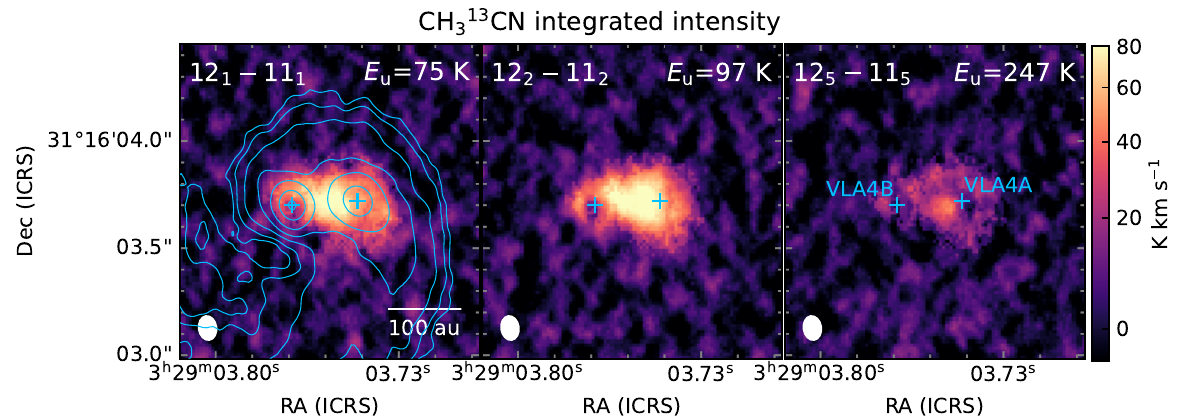}
\end{center}
\caption{Same as Figure \ref{fig:12_mom0} but for the integrated intensity map of the \ce{CH3^{13}CN} J=12-11 with k=1, 2, and 5. The integration range is 5.5-11 km s$^{-1}$. 
}
\label{fig:13_mom0}
\end{figure*}

In Figures \ref{fig:12_mom0} and \ref{fig:13_mom0} we show the integrated intensity maps from three transitions of \ce{CH3CN} and \ce{CH3^{13}CN} that cover different energy levels and suffer less from contamination. For \ce{CH3CN}, it is clear 
that the emission peaks towards VLA 4A while for \ce{CH3^{13}CN}, the emission peaks in the region between the protostars.
In addition, ring-like emission surrounding VLA4B from \ce{CH3CN} and \ce{CH3^{13}CN} is revealed especially for the lines with lower upper energy levels, for which also the emission is the brightest.
Such a ring structure has been found in other sources such as V883 Ori \citep{le19,ya24}.
This can be interpreted
either as due to the high optical depths of the dust or due to destruction of COMs in the inner disk via strong UV or X-ray radiation \citep{ga06,ob09}. An extreme case for this absorption is found in NGC1333 IRAS4A1 with absorption line features \citep{sa19}. 
For SVS13A VLA4B we favor the high dust optical depth scenario because 1) VLA4B indeed has stronger continuum emission than VLA4A, and 2) the UV radiation is likely more powerful in VLA4A as the source is undergoing an accretion burst \citep{hs19}. 

Although the \ce{CH3CN} line emission is dominated by VLA4A, the \ce{CH3^{13}CN} emission peak is located in between VLA4A and VLA4B.
It is unclear at the current stage if this is the material interacting between the binary system \citep{jo22}.
This is consistent with the result from \citep{hs23}; they speculate that \ce{CH3CN} and \ce{CH3^{13}CN} emission comes from different regions or layers due to the different optical depths. With the resolved map, we can clearly see the spatial difference even in the optically thin emission, i.e., K=7 from \ce{CH3CN}.
In other words, these results suggest a spatially inhomogeneous isotopologue ratio of \ce{CH3CN}/\ce{CH3^{13}CN} toward the protobinary SVS13A.

\section{Analysis} \label{sec:analysis}
\subsection{LTE model setup}
We apply local thermal equilibrium (LTE) modeling to the \ce{CH3CN} and \ce{CH3^{13}CN} lines for the whole map (pixel by pixel). A spatial mask is taken using a threshold of S/N>12 from the \ce{CH3CN} K=3 integrated intensity map (Figure \ref{fig:spe_multi}).
To avoid the contamination from other lines, a selected mask at specific velocities is provided manually (such as the blue-shifted side in \ce{CH3CN} K=6 or \ce{CH3^{13}CN} K=3 components).
The LTE model is constructed based on CASSIS and XCLASS manuals \citep{va15,mo17}. The line data are taken from CDMS \citep{mu05,en16} and JPL \citep{pi98}, whereas the original catalogues are from \citep{mu09,mu15}.
The details are in Appendix \ref{app:lte}.

\subsection{Radiative transfer model}
We conduct LTE models throughout the emitting regions pixel by pixel, adopting a one- or two-layer model. Here we introduce these two models and present the model selection in section \ref{sec:clustering}.
\subsubsection{One-layer model} \label{sec:one-layer}
We first conduct a one velocity component LTE modeling which takes into account dust absorption. 
This approach assumes that the dust and gas are co-located instead of assuming the dust as a background to the gas (see \citealt{sa19,ro21,bo21,ya24}), and that the dust and gas temperature are the same, $T_{\rm ex}$. The continuum subtracted spectrum is thus
\begin{equation}
    I_{\nu} (v)= [J_{\nu}(T_{\rm ex})-J_\nu(T_{\rm bg})]  (1-e^{-\tau (v)}) e^{-\tau_{\rm dust}},
\label{eq:rad}
\end{equation}
where $T_{\rm bg}=2.7$ K, $\tau(v)$ is the line optical depth, and $\tau_{\rm dust}$ is the dust optical depth.
The detailed derivation can be found in Appendix \ref{app:lte_one}.
As a result, we have a scale factor of $\exp{(-\tau_{\rm dust})}$. This plays the same role as in \citet{hs23,hs24}, where a beam filling factor was used to scale the emission.
At the edge of the emitting area, a beam might cover a partially line-free region where a beam-filling factor would be needed so that the $\tau$ might be overestimated. We chose a high S/N= 12 as a mask in the integrated intensity map of K$=$3 to avoid this.
We note that an LTE model without considering dust absorption was also tried but failed to interpret the line spectrum: without the scaling factor $e^{-\tau_{\rm dust}}$, the model naturally would underestimate the column density with a small $\tau$ in order to fit the absolute brightness of the observed spectrum, resulting an inconsistent line ratio.

\begin{figure*}
\includegraphics[width=0.99\textwidth]{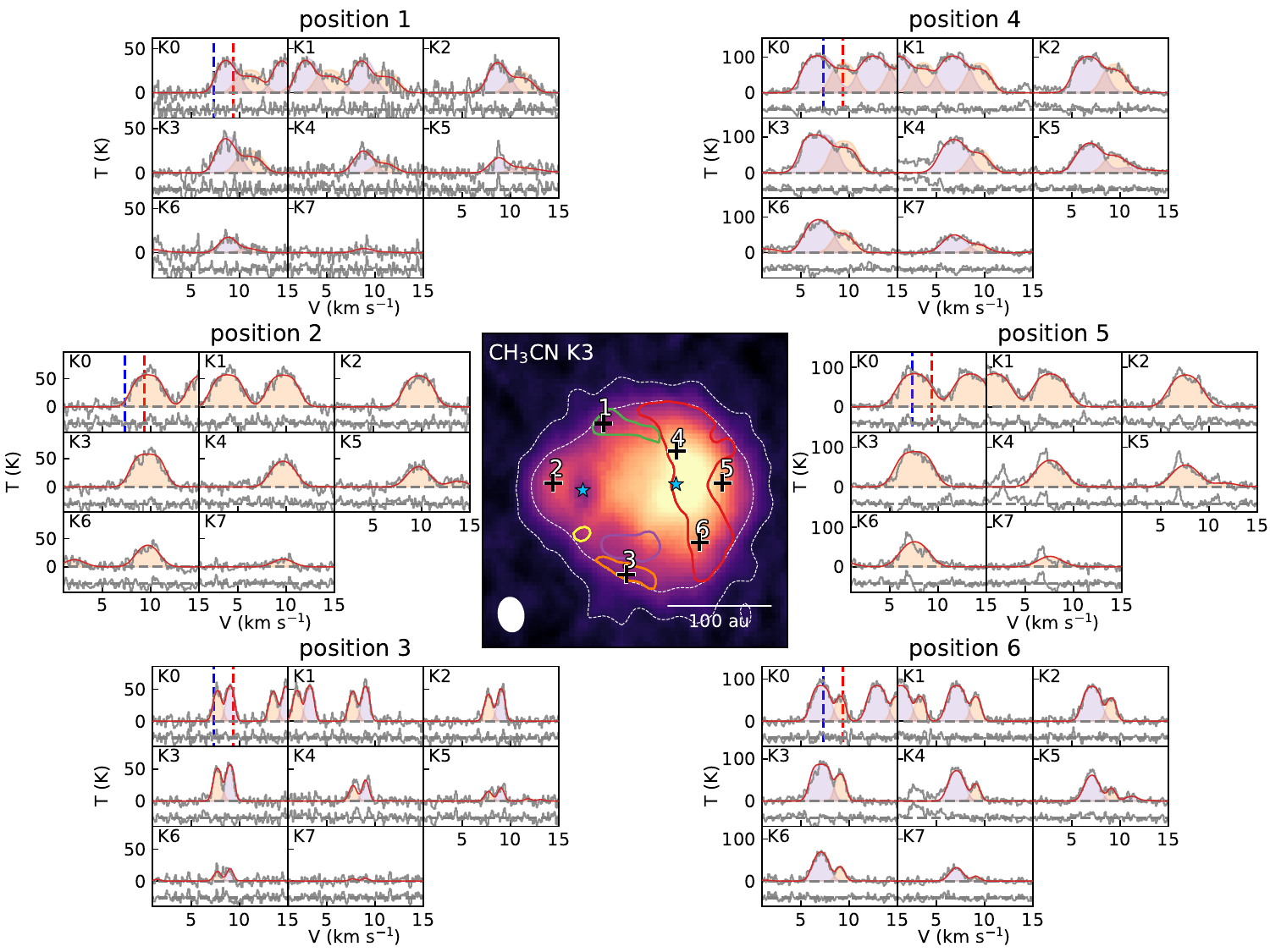}
\caption{Spectra of \ce{CH3CN} toward SVS13A at selected positions marked in the central image of the K=3 integrated intensity map. 
In the central image, the white contours represent the S/N of 3 and 12, as the latter represents the region where LTE fitting is made.
The blue stars indicate the positions of VLA4A and VLA4B.
The kinematic components II-VI (II: red, III: green, IV: orange, V:purple, VI: yellow, see Figure \ref{fig:3d_cluster}) are shown as contours in different colors and for the positions within a contour, the two-layer LTE model is used for fitting (see section \ref{sec:clustering}).
Each spectrum has been divided to eight individual panels centering at the $V_{\rm lsr}$, i.e., 7.36 km s$^{-1}$ for VLA4A and 9.33 km s$^{-1}$ for VLA4B (\citealt{di22} blue and red dashed vertical lines in the K=0 panel), of the K-ladder. The red spectra show the best-fit from the LTE model while the color areas show the 
two-layer model with front (1, orange) and rear (2, purple) layers. 
We note the equation \ref{eq:2_layer} is more complicated, so this colored area only aim to give an idea of $V_{\rm lsr}$ and $\Delta V$.}
\label{fig:spe_multi}
\end{figure*}

\begin{figure*}
\includegraphics[width=0.99\textwidth]{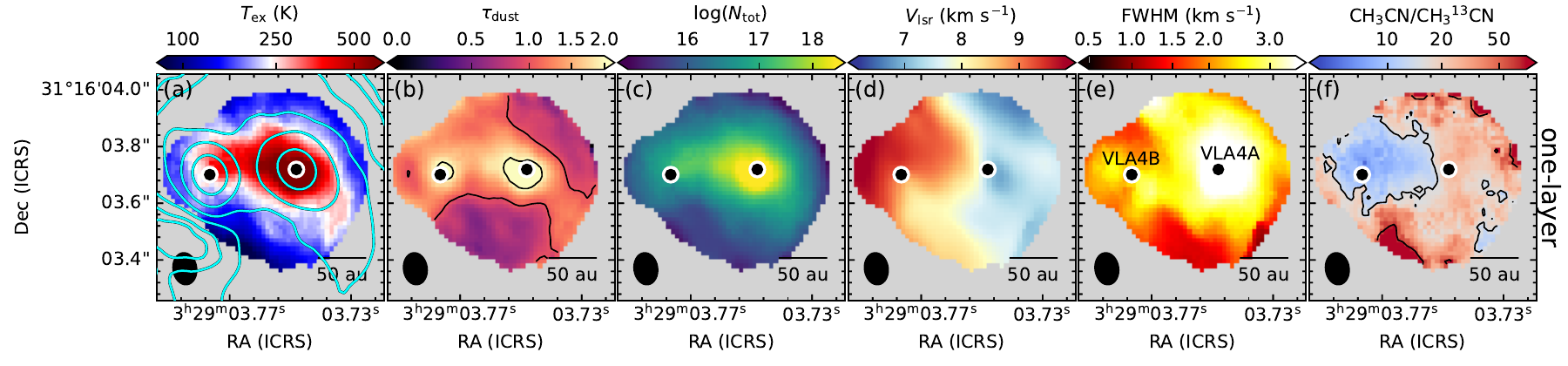}
\caption{Fitting results of the one-layer LTE model. Each panel shows a map of the free parameter with the label of the colorbar on the top. The contours on the first panel represents the 1.3 mm continuum emission (Figure \ref{fig:12_mom0}). The contours of the 2nd column represents  $\tau$ at 1 and 2, and those on the 6th column show the isotopologue ratios of 15 and 68. In the 4th column showing $V_{\rm lsr}$ map, the systemic velocities of VLA4A and VLA4B are 7.36 km s$^{-1}$ and 9.33 km s$^{-1}$, respectively \citep{di22}.}
\label{fig:para_one}
\end{figure*}

The spectra in Figures \ref{fig:spe_vla} (position VLA4B) and \ref{fig:spe_multi} (positions 2 and 5) show the one-layer model fit toward these positions.
Figure \ref{fig:para_one} shows the maps of the six free parameters from the fit. The temperature peaks toward VLA4A. We note that the partition function $Q(T)$ is provided by CDMS up to 900 K, so the boundary of the fitting is set to 800 K. 
The derived $\tau_{\rm dust}$ from the line fit is quite similar to the continuum emission, supporting that the dust optical depths are reasonably estimated. The peak $\tau\sim2.5$ suggests that only $\sim10$\% of line emission might escape from these regions. There are also regions with $\tau<1$, suggesting that the continuum emission is not optically thick through the entire line emitting region. Thus, it is reasonable to assume coupling of gas and dust rather than an optically thick continuum emission as background. \ce{CH3CN} column density is higher toward VLA4A as expected. The $V_{\rm lsr}$ and FWHM maps are in general similar to the average-velocity maps and moment 2 maps from previous high-resolution COM maps \citep{di22,bi22}.
The isotopologue ratio \ce{CH3CN}/\ce{CH3^{13}CN} is taken as $\frac{N_{\rm tot,\ce{CH3CN}}}{N_{\rm tot,\ce{CH3^{13}CN}}}$ (see equations \ref{eq:tau_v} and \ref{eq:tau}). As a result, the spatially inhomogeneous distribution is presented.

\subsubsection{Two-layer model} \label{sec:two-layers}
Although the LTE model with dust absorption can interpret the relative intensities of the K-ladder transitions, the high spectral resolution data reveal double-peak features in some areas (Figure \ref{fig:spe_multi}). \citet{hs24} have conducted multi-component LTE fitting of O-bearing COMs toward SVS13A, decomposing different kinematic components.
In that work, the kinematic components are assumed spatially separated within the NOEMA beam.
However, with the spatially-resolved ALMA map, the double peak spectral profiles are still present in particular regions, indicating two components along the line of sight. Thus, we conduct the two-layer LTE model to interpret these spectra. Each layer has its own excitation temperature ($T_{\rm ex}$), dust optical depth ($\tau_{\rm dust}$), \ce{CH3CN} column density ($N_{\rm tot}$), central velocity ($V_{\rm LSR}$), linewidth ($\Delta V$), and isotopologue ratio (\ce{CH3CN}/\ce{CH3^{13}CN}); twelve free parameters are included. The continuum subtracted spectrum is
\begin{equation}\label{eq:2_layer}
\begin{aligned}
    I_{\nu} (v) = 
        &\, [J_{\nu}(T_{\rm ex, 1})-J_\nu(T_{\rm ex, 2})]  (1-e^{-\tau_1 (v)}) e^{-\tau_{\rm dust, 1}} \\
        +&\, [J_{\nu}(T_{\rm ex, 2})-J_\nu(T_{\rm bg})]  (1-e^{-[\tau_1 (v)+\tau_2 (v)]}) e^{-(\tau_{\rm dust, 1}+\tau_{\rm dust, 2})},
\end{aligned}
\end{equation}
and the detailed derivation is in Appendix \ref{app:lte_two}.
This equation can be considered as a variation from the two-layer model in \citet{my96} and its family \citep{le01,di01,de05,ka12}. The original two-layer model is applied to infall motion at molecular core where the dust absorption might be negligible. Equation \ref{eq:2_layer} will converge to the original two-layer model when $\tau_{\rm dust,1}=0$ and $\tau_{\rm dust,2}=0$. In addition, the original two-layer model assumes two uniform layers with the same optical depth $\tau(v)$ (see equation \ref{eq:tau_v}) as symmetrically infalling. Here the LTE model requires number densities of population for energy levels (defining $\tau$) to follow the Boltzmann distribution with the given $T_{\rm ex}$ (equations \ref{eq:tau} and \ref{eq:boltzmann}). We find that a symmetric model, i.e., the same velocity dispersion $\sigma$ and $N_{\rm tot}$ for the two layers, is insufficient for some regions, for example, position 4 in Figure \ref{fig:spe_multi} with linewidths of 0.9 and 1.5 km s$^{-1}$ for the first and the second layer, respectively.
It is noteworthy that this 2-layer model will also converge to the gas-dust decoupling model when $\tau_{\rm dust,1}=0$ (gas front layer) and $\tau_2(v)=0$ (dust rear layer) as \citet{ro21}. Clearly, if we have an optically thick dust continuum as the rear layer ($\tau_{\rm dust,2}\gg1$), only the first term in equation \ref{eq:2_layer} is left with $T_{\rm ex,2}$ representing the background temperature of the dust layer.

We perform the fit throughout the line emitting region (Figure \ref{fig:spe_multi}). Such a fit with many free parameters is difficult to converge and suffers from degeneracy (see Appendix \ref{app:degeneracy}). To address these issues, good initial guesses are required. We first perform Markov Chain Monte Carlo (MCMC) sampling of a grid of positions in the emitting regions (in total 41 positions with gaps around emission of roughly beam size). 
We tune the sampling boundary (prior range) of the MCMC fitting and find the most reasonable solution. 
For example, the fitting results should not show dramatic difference in between two neighboring positions.
These solutions at each points in the grid are then interpolated to a map of initial guesses for the fitting.
Figure \ref{fig:spe_multi} shows the spectra of a few selected positions for which the best-fit from MCMC sampling are plotted (with one- or two-layer model in use, see section \ref{sec:clustering}).

\begin{figure}
\begin{center}
\includegraphics[width=0.5\textwidth, trim=70 40 30 70, clip]{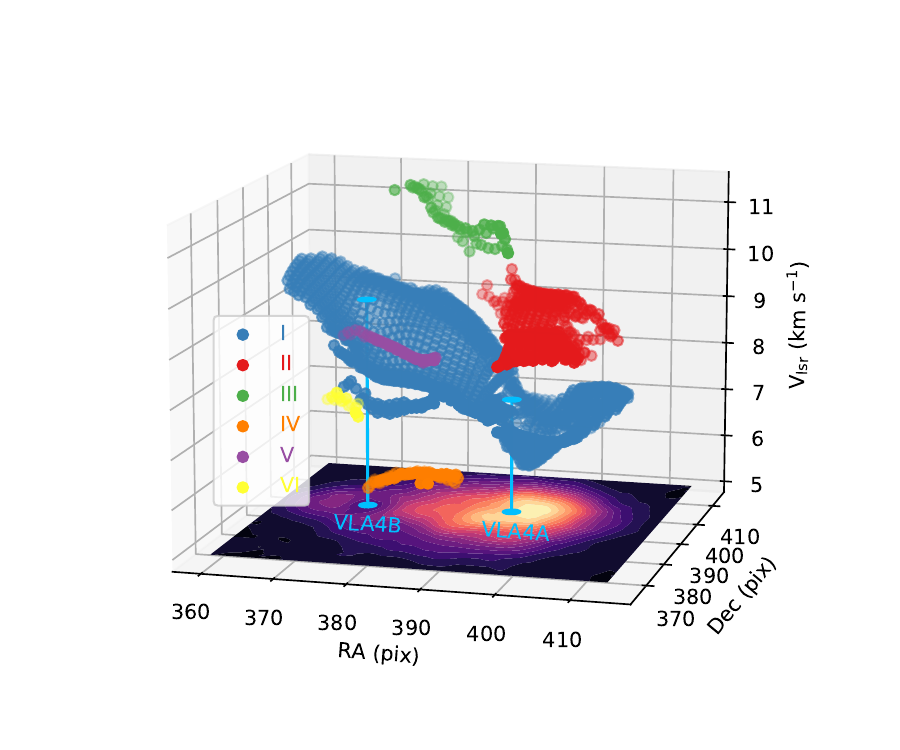}
\end{center}
\caption{PPV diagram from the combination of one- and two-layer model fitting. The bottom x-y image is the \ce{CH3CN} J=12-11 K=3 integrated intensity map. Above it, each position can have one or two velocities. The color of data points indicates the kinematic component from clustering.
The two cyan vertical lines indicates the positions and the systemic velocities of VLA4A (7.36 km s$^{-1}$) and VLA4B (9.33 km s$^{-1}$).
}
\label{fig:3d_cluster}
\end{figure}

\begin{figure}[t]
\includegraphics[width=0.5\textwidth]{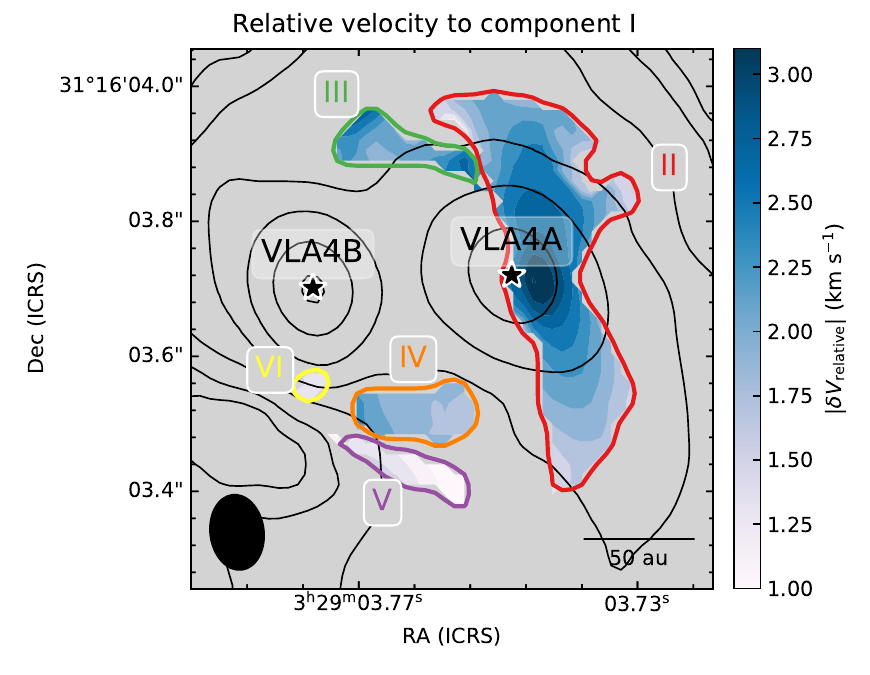}
\caption{Absolute relative velocity of the component II-VI with respect to the main component I. The contours represent the continuum emission with the same levels as Figure \ref{fig:spe_vla}. The color contours show the area of component II-VI.}
\label{fig:infall}
\end{figure}

\subsection{Clustering into kinematic components}\label{sec:clustering}
Here we discuss how we select the one- or two-layer model in each pixel and cluster them into kinematic components. 
We use a criterion for the two layer-model: the difference of the centroid velocities ($V_{\rm in}=V_{\rm LSR,1}-V_{\rm LSR,2}$) is larger than the velocity dispersions of both layers, $\sigma_1$ and $\sigma_2$.
This threshold set a strict situation for using two-layer model only when the double-peak is clearly seen in the position.
For example, in Figure \ref{fig:spe_multi} position 4 and 5 using one-layer model, the spectral profiles in fact shows blue or red asymmetry which is considered as an indicator of infall or expansion motions in cloud cores \citep{ma97,ta98,le01,em09,pi12,hs15,re22}.
Although the two-layer model better reproduces the spectra profile with lower reduced $\chi^2$ and Akaike information criterion (AIC, see appendix \ref{app:degeneracy}), the one-layer model is chosen in case where no clear double peak is presented. It is possible that the two layers are blended with similar centroid velocities or broad linewidths. For example in Figure 7 in \citealt{bi17} with \ce{CH3OH} isotopologues toward SVS13A, a broken power law is seen in the rotation diagram, decomposing components without spatially or spectrally resolved data. However, in our case with only eight energy levels from \ce{CH3CN} J=12-11 K=0-7 (in the rotational diagram), we decide to be more conservative, using one-layer model for positions without a clear double peak. 
More importantly, to study the kinematic structure, these regions with small relative velocity between two layers (see also Figure \ref{fig:aic} right panel) will not be decomposed or extracted as a new kinematic component by the clustering algorithms in the position-position diagram.

\begin{figure*}
\includegraphics[width=0.99\textwidth]{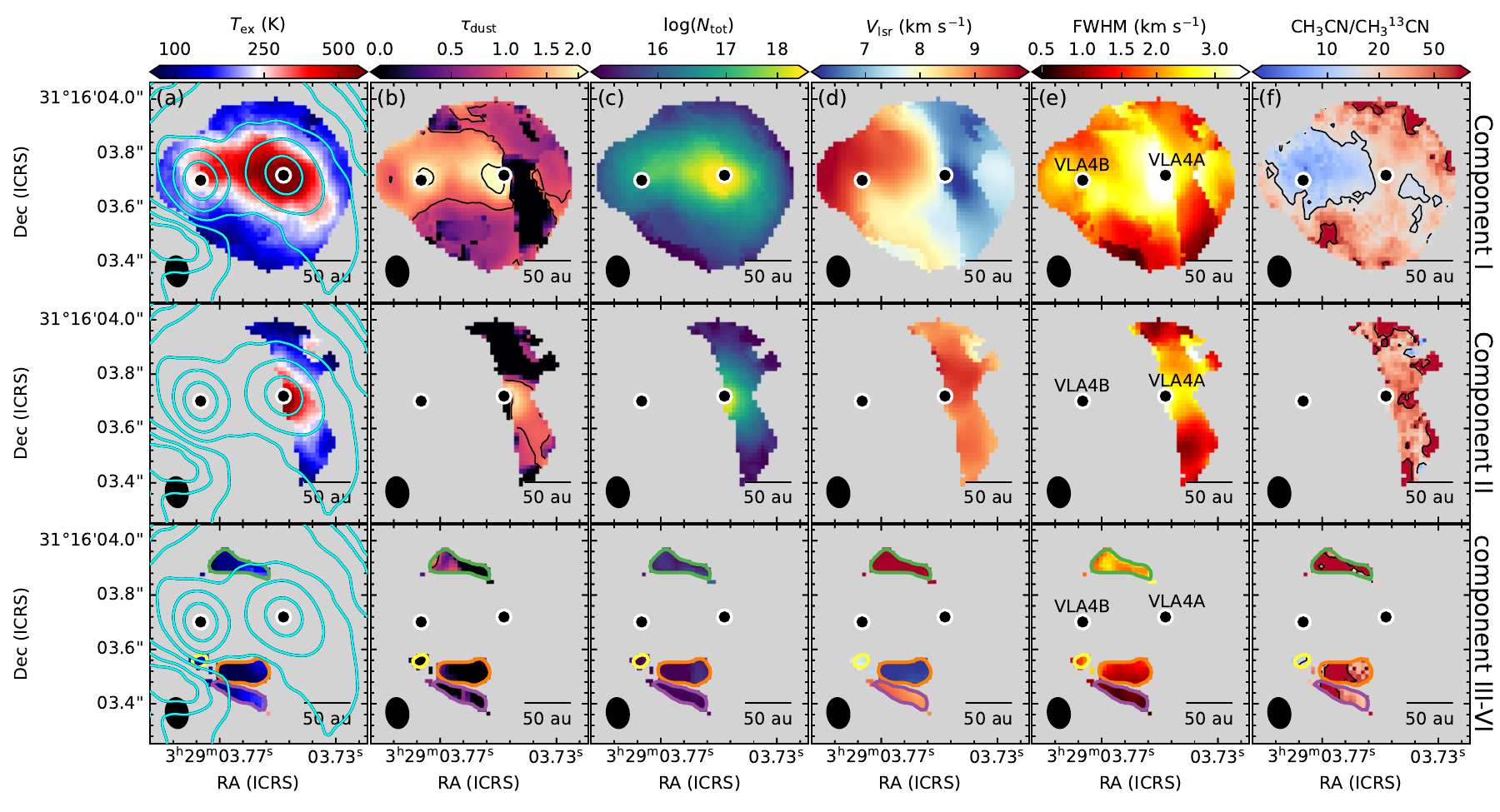}
\caption{Same as Figure \ref{fig:para_one} but for the clustering components. Note the discarded points are attributed to one-layer fitting in component I. The top, middle, and bottom shows the parameters for component I, II, and that including III-VI. In the 4th column showing $V_{\rm lsr}$ map, the systemic velocities of VLA4A and VLA4B are 7.36 km s$^{-1}$ and 9.33 km s$^{-1}$, respectively \citep{di22}.}
\label{fig:para_multi}
\end{figure*}

After the model selection, we have a final map at each position containing one or two velocity components (from the one- or two-layer model respectively). Figure \ref{fig:3d_cluster} shows the position-position-velocity diagram (PPV), because it can better decompose the \ce{CH3CN} distribution than P-V diagrams. We use the clustering algorithm DBSCAN \citep{es96,va23,gi24} with {\tt eps=0.3} and {\tt min\_samples=10}, clustering these points into six groups. For discarded points, we simply use the one-layer model at such positions, so that they go back to the major kinematic component I (Figure \ref{fig:3d_cluster}). 
As a result, two major kinematic components I and II, and four small components III-VI are identified (Figure \ref{fig:3d_cluster}). Table \ref{tab:para} lists their physical parameters. 
Because the two-layer model in fact determines the front layer and the rear layer, i.e., equation \ref{eq:2_layer}, the infall or expansion of the two layers can be obtained.
A positive value of $V_{\rm in}$ in Table \ref{tab:para} indicate expanding motion while a negative one indicates an infalling motion \citep{my96,ma97}.
However, there is a caveat to determine this given the degeneracy of the model. For example, although the double peak profiles with blue excess are clearly seen toward the region with component II as infalling signatures, we can find another fit for expanding motion due to the asymmetric layers (see Appendix \ref{app:degeneracy} for the degeneracy).
To completely break the degeneracy, multi-band observations would help with measuring dust opacity at different wavelengths (for example, optically thick dust emission as \citealt{pi12}).
Finally the absolute value of the relative velocity of each component with respect to component I is presented in Figure \ref{fig:infall}. Figure \ref{fig:para_multi} shows the best-fit parameters of all six components and Table \ref{tab:para} lists them.

\begin{table*}
    \centering
    \caption{Kinematic components}
    \begin{tabular}{ccccccccc}
    \hline\hline
& size
& $T_{\rm ex}$
& $\tau_{\rm dust}$
& log$_{10}(N_{\rm tot})$
& $V_{\rm LSR}$
& $V_{\rm in}$ $^a$
& $\Delta V$
& \ce{CH3CN}/\ce{CH3^{13}CN} \\
component
& (au$^2$)
& (K)
& 
& (cm$^{-2}$)
& (km s$^{-1}$)
& (km s$^{-1}$)
& (km s$^{-1}$)
& \\
\hline
one-layer       & 27660 & 210.6$^{+72.7}_{-42.1}$       & 1.1$^{+0.3}_{-0.3}$   & 16.4$^{+0.6}_{-0.4}$  & 8.0$^{+0.8}_{-0.6}$     & -     & 1.0$^{+0.2}_{-0.2}$   & 19.8$^{+6.9}_{-5.9}$\\
\hline
I       & 27660 & 205.5$^{+79.1}_{-43.1}$       & 0.9$^{+0.4}_{-0.3}$   & 16.3$^{+0.5}_{-0.4}$  & 7.9$^{+0.8}_{-0.7}$   & -       & 0.9$^{+0.2}_{-0.2}$   & 19.5$^{+8.6}_{-5.6}$\\
II      & 7282  & 178.9$^{+73.7}_{-47.0}$       & 0.4$^{+0.7}_{-0.4}$   & 15.8$^{+0.6}_{-0.3}$  & 9.0$^{+0.2}_{-0.1}$   & -2.1$^{+0.3}_{-0.3}$    & 0.8$^{+0.2}_{-0.2}$   & 38.6$^{+34.0}_{-11.9}$\\
III     & 1139  & 105.8$^{+38.6}_{-6.6}$        & 0.4$^{+0.3}_{-0.4}$   & 15.6$^{+0.1}_{-0.1}$  & 10.6$^{+0.4}_{-0.2}$  & -2.2$^{+0.1}_{-0.2}$    & 0.9$^{+0.1}_{-0.1}$   & $>$80.0\\
IV      & 1274  & 77.6$^{+31.5}_{-24.7}$        & 0.0$^{+0.1}_{-0.0}$   & 15.3$^{+0.3}_{-0.2}$  & 6.2$^{+0.1}_{-0.1}$   & 1.9$^{+0.1}_{-0.1}$     & 0.5$^{+0.1}_{-0.1}$   & 68.5$^{+11.5}_{-38.3}$\\
V       & 869   & 139.9$^{+18.2}_{-33.8}$       & 0.0$^{+0.3}_{-0.0}$   & 15.3$^{+0.0}_{-0.1}$  & 9.0$^{+0.1}_{-0.1}$   & 1.1$^{+0.2}_{-0.2}$     & 0.3$^{+0.0}_{-0.0}$   & 42.5$^{+37.5}_{-15.1}$\\
VI      & 270   & 121.1$^{+13.9}_{-11.8}$       & 0.0$^{+0.0}_{-0.0}$   & 15.3$^{+0.1}_{-0.1}$  & 7.5$^{+0.1}_{-0.1}$   & 1.3$^{+0.0}_{-0.1}$     & 0.8$^{+0.1}_{-0.1}$   & 15.0$^{+5.9}_{-3.4}$\\
\hline
    \end{tabular}\\
\tablefoot{The numbers in the upper and lower limits are taken from the distributions in quantiles of 0.25 and 0.75.\\
\tablefoottext{a}{The relative velocity of the component related to the main component I. A positive number indicates expanding motion and a negative one indicates infalling but should be taken with caution, see appendix \ref{app:degeneracy}.}
}
\label{tab:para}
\end{table*}

\begin{figure*}[t]
\begin{center}
\includegraphics[width=0.7\textwidth]{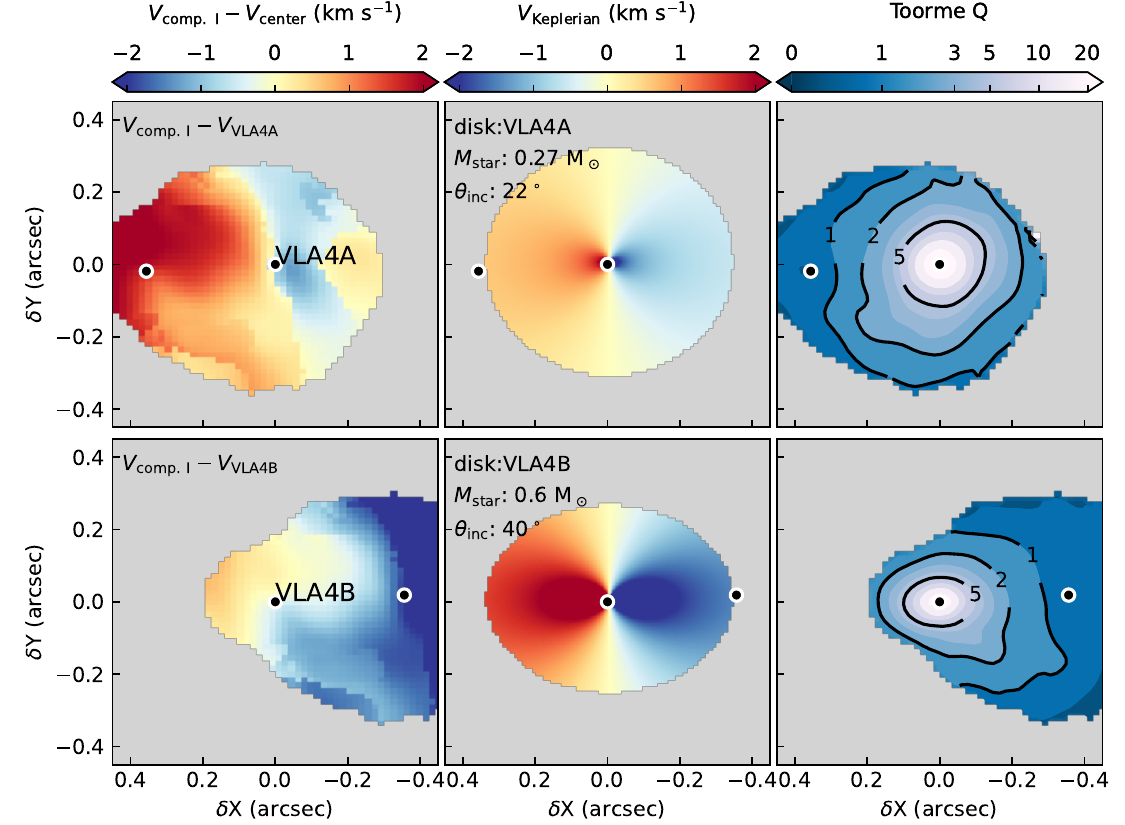}
\end{center}
\caption{(Left column) Velocity map of component I (Figure \ref{fig:para_multi}) relative to VLA4A (top, $V_{\rm VLA4A}=7.36$ km s$^{-1}$) and VLA4B (bottom, $V_{\rm VLA4B}=9.33$ km s$^{-1}$). (Middle) Keplerian disk model using the $M_{\rm star}$ and $\theta_{\rm inc}$ from \citet{di22}. We note that here the inclination angle of VLA4B is from the dust continuum as the gas disk is not seen. (Right) Toomre Q map derived using the physical parameters from \ce{CH3CN} plus models of rotation from the middle panel.}
\label{fig:Q}
\end{figure*}

\section{Discussion}\label{sec:discussion}
\subsection{Kinematic structures}\label{sec:kin}
The velocity map, panel (d) of Figure \ref{fig:para_one}, in general reveals similar structures to average-velocity maps from literature \citep{di22,bi22}.
However, our two-layer model aims at ambitiously studying the motion in the line of sight of the system (Figure \ref{fig:para_multi}).
The velocity gradient along west-east of the major component I is larger than that in the one-layer model because at the west side, the red-shifted emission is decomposed to the component II.
Figure \ref{fig:infall} shows the absolute value of the  relative velocity $|\delta V_{\rm relative}|$ of these components with respect to the main component I.
A major portion of Component II is around VLA4A, perhaps also affected or induced by the dusty spiral which is considered as a possible infalling streamer \citep{hs23}. 

Figure \ref{fig:infall} reveals a velocity pattern  ($|\delta V_{\rm relative}|$) of component II relative to I with a radial symmetric structure centered to the west of VLA4A. We note again that, although the two-layer model has defined front and rear layers with different velocities, these two layers can frequently be hard to disentangle (see Appendix \ref{app:degeneracy}). Thus, we are not able to determine if it is an infall or outward motion only from the spectra.
However, the majority of component II shows blue excess favor infall motion. In addition, the radial symmetric structure in Figure \ref{fig:infall}, i.e., the relative velocity between component I and II, implies this traces an infall motion centering around VLA4A.
If this is the case, it might help understanding the nature of the possible infalling streamer \citep{hs23}. In this case, the secondary source VLA4A 
is fed by the infalling streamer during an accretion outburst.
A caveat is that, the possible infalling streamer found by \citet{hs23} via DCN has a landing velocity at the blue-shifted emission $7-8$ km s$^{-1}$ similar to \ce{C^{18}O} from the spiral \citep{to18}. This is different from the component II with velocity around 9 km s$^{-1}$ (Table \ref{tab:para} and Figure \ref{fig:para_multi}). This implies that the component II from the \ce{CH3CN} emission is not directly associated with the dusty spiral.

Regarding the rotation, \citet{di22} found a disk surrounding VLA4A and a circumbinary disk via ethylene glycol and CS observations, respectively. The $V_{\rm lsr}$ maps from our LTE fitting, i.e., panel (d) in Figures \ref{fig:para_one}, are broadly similar to the mean-velocity map in \citet{di22}. The S-shape pattern in between the blue- and red-shifted emission can be explained by the simultaneous presence of infall and rotation \citep{fl23}. 
We looked at position-velocity (PV) diagrams (Figures \ref{fig:12_pv} and \ref{fig:13_pv}), where the position axis goes from west to east and passes through A and B.
In the PV-diagrams, likely only the blue-shifted emission in the west of the Keplerian rotation is seen via \ce{CH3CN} and \ce{CH3^{13}CN}, as in component I (Figure \ref{fig:para_multi}). Figure \ref{fig:Q} compares the velocity pattern of component I with thin disk models given the parameters from \citet{di22}. Consistent with the literature, the model can broadly describe the VLA4A disk while in VLA4B there is no detection of a gas disk (see Figures \ref{fig:12_pv} and \ref{fig:13_pv}), for which the inclination angle is obtained from dust continuum. 
This makes component II interesting in that it likely is associated with infalling materials, but not directly to the infalling streamer or the continuum spiral.

\subsection{Physical conditions}
Figures \ref{fig:para_one} and \ref{fig:para_multi} show the gas temperature structures from the \ce{CH3CN} K-ladder. The temperature map of the major component I is similar to that of the one-layer model. Both components I and II have a peak temperature toward VLA4A.
This could also explain the high column density of \ce{CH3CN} at VLA4A, where molecules are sublimated from the dust mantles.
The component II is relatively cooler than component I (Table \ref{tab:para}). As the component I is likely associated with the rotating disk or envelope around VLA4A (Figures \ref{fig:3d_cluster} and \ref{fig:12_pv}), the component II might trace materials induced by the possible infalling streamer. 

Radiation from central protostars is usually considered as the main heating source, which is used to determine the temperature from previous modeling works \citep{ra17,hs18,hs19,mu22a}.
Toward SVS13A, although the temperature is mostly determined by the protostellar radiation, it is unlikely the only factor in order to reproduce the asymmetric structure. 
First, toward VLA4B, the gas temperature in its west side is clearly higher than that in the east. Likely, the circumbinary material in between VLA4A and VLA4B is relatively warmer. More interestingly, the temperature is relatively high along the spiral, the possible infalling streamer (Figures \ref{fig:para_one} and \ref{fig:para_multi}). Mechanical heating has been suggested to interpret such asymmetric temperature structures as in IRAS16293-2422 \citep{ma22}. If the dusty spiral traces a streamer accreting onto the protostellar or circumbinary disk, it can induce shocked gas and heat up the material. For example, shocked gas traced via SO has been detected toward the landing position of an infalling streamer in HL Tau \cite{ga22} and in Per-emb-50 \cite{va22}.
High-spatial and high-spectral resolution observations of SO might help disentangle the shocked gas induced in SVS13A.

\subsection{Toomre Q analysis}
Disk fragmentation is considered as a formation process of close binary systems \citep{of23}. The Toomre Q parameter is widely used to test if a disk is gravitationally unstable as
\begin{equation}\label{eq:toomre}
Q=\frac{c_S \Omega}{\pi G \Sigma},
\end{equation}
where $c_S$ is the sound speed, $\Omega$ is the angular velocity, $G$ is the gravitational constant, and $\Sigma$ is the disk surface density \citep{to64}. Here we use the $T_{\rm ex}$ from the one-layer model to derive the sound speed (Figure \ref{fig:para_one}, see also \citealt{ah23}). We use $\tau$ map (Figure \ref{fig:para_one}) to estimate $\Sigma$ assuming the dust opacity $\kappa=0.899$ cm$^{2}$ g$^{-1}$ at 1.3 mm \citep{os94,to18}, a gas to dust mass ratio of 100, with inclination correction from \citet{di22}. It is noteworthy that the dust opacity can bring an uncertainty of a few \citep{ti18}.
To estimate the angular velocity, we compare the velocity map from Figure \ref{fig:para_multi} and thin disk models for VLA4A and VLA4B given the star masses and inclination angle from \citet{di22} (Figure \ref{fig:Q}). We note here the component I in Figure \ref{fig:para_multi} is used because the one-layer velocity map (Figure \ref{fig:para_one}) includes the contribution from component II with the red-shifted emission in the east (Figures \ref{fig:para_multi} and \ref{fig:12_pv}). By assuming Keplerian rotation, we obtain $Q$ maps for VLA4A and VLA4B (Figure \ref{fig:Q}).

The Q parameter maps show that the disks are in general stable near the protostars (Q $>1$). However, VLA4A is likely undergoing an accretion outburst, suggesting that the high temperature is temporary. If we assume $T\propto L^{0.2}$ \citep{go74} and the luminosity was 100 times less during the quiescent phase, the $c_s$ as well as the $Q$ parameter should be a factor of $\sim$0.6 of the current values. This makes more area in the disk gravitationally unstable so that the disk fragmentation might occur during the quiescent phase causing the binary formation or triggering the accretion burst \citep{bo08,vo05,vo10,of09,ma11}. Also, surrounding VLA4A is the potential infalling pattern (Figure \ref{fig:infall}) and the possible infalling streamer \citet{hs23} which could be in part responsible for the accretion burst via gravitational instability \citep{pi23}.

\subsection{\ce{CH3CN}/\ce{CH3^{13}CN} spatial distribution}

The \ce{^{12}C}/\ce{^{13}C} ratio has been measured toward different protostellar systems \citet{bu24} in prep., and
is usually found to be lower relative to the canonical ratio in the local ISM \ce{^{12}C}/\ce{^{13}C}$\simeq$68 \citep{mi05}.
It is also found to vary at different distances from the central source \citep{yo22,yo24,be24}
Toward SVS13A, \citet{hs23} find a low value of 16 as they suggest that the \ce{CH3CN} and \ce{CH3^{13}CN} might trace different regions or layers toward SVS13A. 
This is supported by the integrated intensity maps of Figures \ref{fig:12_mom0} and \ref{fig:13_mom0}. In other words, the isotopologue ratios can vary across SVS13A. 
With the LTE model (Figures \ref{fig:para_one} and \ref{fig:para_multi}), we find low values in comparison with the canonical one throughout the maps except for component III and IV, but with large error (Table \ref{tab:para}). Other than these two components, the \ce{CH3CN}/\ce{CH3^{13}CN} ratios $\sim$$20-40$ are in general lower than the canonical value of 68. Interestingly, it is even lower around the primary source VLA4B from both one- and two-layer model (Figures \ref{fig:para_one} and \ref{fig:para_multi}). This suggests a chemical segregation between VLA4A and VLA4B as reported by \citet{bi22} with other COMs. Furthermore, as the secondary VLA4A is likely connected to the dusty spiral, possibly tracing an infalling streamer \citep{hs23}, this hints that the infalling streamer might affect the chemistry in the inner region. For example, via \ce{H3CN}, an infalling streamer is suggested to deliver chemical fresh materials from the cores into the disks in Per-emb-2 \citep{pi20}.

\section{Summary}\label{sec:summary}
We present \ce{CH3CN} and \ce{CH3^{13}CN} observations toward SVS13A using ALMA with a spatial resolution of $\sim23\times32$ au and spectral resolution of $\sim$0.08 km s$^{-1}$. We conduct LTE radiative transfer fitting of line emission. Our main findings are summarized:
\begin{enumerate}
\item The \ce{CH3CN} and \ce{CH3^{13}CN} integrated intensity are tracing material near both protostars as well as material along the direction connecting both protostars. However, the \ce{CH3CN} is the brightest towards the secondary protostar VLA4A while its \ce{^{13}C} isotopologue is the brightest in between the two protostars.  In addition, towards the primary VLA4B, the emission from both \ce{CH3CN} and \ce{CH3^{13}CN} shows a dip at the location of the continuum peak resulting in a ring-like morphology. This can be interpreted as dust absorption.

\item We modeled the lines from \ce{CH3CN}/\ce{CH3^{13}CN} considering the optical depth of the dust (i.e, dust absorption), assuming that the dust and gas are co-spatial and have the same temperature. We inferred temperatures $\ga$100 K in the line-emitting region, with the highest temperatures towards VLA4A (the secondary, $>$500 K), the region between the protostars ($\sim$300 K), and that along the possible infalling streamer ($\sim$250 K). The inhomogenous distribution of the temperature suggests that in addition to irradiation from the outbursting protostar VLA4A, there is also heating from shocks possibly induced by an infalling streamer.

\item Toward some regions with resolved double-peak line profiles, we conduct two-layer LTE radiative transfer modeling to reproduce the features. This enables us to identify two major kinematic components I and II. While the component I is likely associated with rotating disk or envelope around VLA4A and VLA4B, the component II might be associated with the dusty spiral structure, a possible infalling streamer, around VLA4A. 
The velocities of this material differs from the main component by 1 to 3 km s$^{-1}$ and could be connected to the infalling material towards the secondary VLA4A which is undergoing an accretion burst.


\item Using the individual protostellar masses, the measured gas temperature for \ce{CH3CN} and dust optical depth at 1.3 mm, we derive maps of the Toomre Q parameter around the protostars. 
We find broadly gravitationally stable disks (or rotating materials) with the current conditions.  However, in the past, previous to the accretion outburst conditions experiences by VLA4A, the disk was cooled down and massive enough in some parts being gravitationally unstable. It could have fragmented to trigger the binary formation or accretion bursts.

\item The \ce{CH3CN}/\ce{CH3^{13}CN} ratio is lower than the canonical \ce{^{12}C}/\ce{^{13}C} ratio in the ISM over the whole region, but with variations across the map. The ratio towards the primary VLA4B, as well as for the material between the protostars, is around 10. On the other hand, the western part of the emission which includes VLA4A shows values between 15 and 50. Streaming material towards VLA4A could be the origin of this difference. 

\end{enumerate}

\begin{acknowledgements}
We thank Dr. Naomi Hirano for the insight discussion.
T.-H. H., J.E.P., P.C., M.T.V, and M.J.M. acknowledge the support by the Max Planck Society.
T.-H. H. is supported by Taiwan Astronomical Research Alliance (TARA) with National Science and Technology Council (NSTC 113-2740-M-008-005). TARA is committed to advancing astronomy in Taiwan and paving the way for the establishment of a national observatory.
This paper makes use of the following ALMA data: ADS/JAO.ALMA\#2022.1.00479.S. ALMA is a partnership of ESO (representing its member states), NSF (USA) and NINS (Japan), together with NRC (Canada), NSTC and ASIAA (Taiwan), and KASI (Republic of Korea), in cooperation with the Republic of Chile. The Joint ALMA Observatory is operated by ESO, AUI/NRAO and NAOJ.

\end{acknowledgements}

\bibliographystyle{aa}
\bibliography{SVS13A}


{\it Software:} Numpy (\citealt{numpy}),
Scipy (\citealt{scipy}), 
APLpy (\citealt{aplpy}),
Matplotlib (\citealt{matplotlib}), 
Astropy (\citealt{astropy})
CASA (\citealt{casa})

\clearpage
\onecolumn

\appendix

\section{LTE Radiative transfer models}
\subsection{Boltzmann distribution and optical depth}\label{app:lte}
LTE-radiative transfer model are constructed to fit the line profiles of the \ce{CH3CN} J=12-11, K=0-7 and \ce{CH3^{13}CN} J=12-11, K=0-5 transitions given the observed wavelength coverage. 
This is done following the XCLASS and CASSIS manuals\footnote{http://cassis.irap.omp.eu/docs/RadiativeTransfer.pdf} \citep{mo17} and \citep{va15}.
The LTE $\tau (v)$ as a Gaussian function of velocity ($v$) was first constructed:
\begin{equation}
    \tau (v)=\sum_{\rm i} \tau_{\rm i,peak} \exp(-\frac{(v-V_{\rm LSR})^2}{2\sigma^2}),
    \label{eq:tau_v}
\end{equation}
where $i$ is the transition including \ce{CH3CN} and \ce{CH3^{13}CN}, $\sigma$ is the velocity dispersion ($\sigma\sim\frac{\Delta V}{2\sqrt{2\ln{2}}}$), $V_{\rm LSR}$ is the centroid velocity, 
\begin{equation}
    \tau_{\rm peak}=\frac{A_{\rm ul} c^3}{8\pi \nu^3 \Delta V \frac{(\pi ln2)^{1/2}}{2}} N_{\rm u} (e^{h\nu/kT_{\rm ex}}-1),
    \label{eq:tau}
\end{equation}
for which $A_{\rm ul}$ is the Einstein coefficient, c is light speed, and $\nu$ is the rest frequency of the transition from the Cologne Database for Molecular Spectroscopy (CDMS\footnote{https://cdms.astro.uni-koeln.de/cdms/portal/}; \citealt{mu05,en16}).
$N_{\rm u}$ is the column density at the upper energy state as
\begin{equation}\label{eq:boltzmann}
    N_{\rm u}=N_{\rm tot}\frac{g_{\rm u}}{Q(T_{\rm ex})} e^{-E_{\rm u}/kT_{\rm ex}},\qquad
    g_{\rm u} =
    \begin{cases} 
    50 & \text{if } K = 0, 1, 2 \ldots \\
    100 & \text{if } K=3n
    \end{cases}
\end{equation}
where $g_{\rm u}$ is the degeneracy, $Q(T_{\rm ex})$ is the partition function, and $E_{\rm u}$ is the upper energy level.
We note here the partition function $Q(T)$ of \ce{CH3^{13}CN} is taken from \ce{CH3CN} due to the lack of the vibrational correction. This causes at most $0.06\%$ error at $T<500$ K (\citealt{hs23}).

\subsection{Radiative transfer model with dust optical depth}\label{app:lte_one}
Here we discuss the contribution of dust absorption in the LTE radiative transfer model.
Under the LTE assumption with the same gas temperature and dust temperature ($T_{\rm ex}=T_{\rm dust}$), the gas and dust are assumed to share the same space. The continuum-subtracted line emission with dust absorption can be expressed as:
\begin{equation}
    \begin{aligned}
    \Delta I_{\nu} (v) 
        = &\, I_{\rm line} (v) - I_{\rm dust}    \\
        = &\, [J_{\nu}(T_{\rm bg})e^{-[\tau (v)+\tau_{\rm dust}]} + J_{\nu}(T_{\rm ex})(1-e^{-[\tau (v)+\tau_{\rm dust}]})] \\
        &\, - [J_{\nu}(T_{\rm bg})e^{-\tau_{\rm dust}} + J_{\nu}(T_{\rm ex})(1-e^{-\tau_{\rm dust}})] \\
        = &\, [J_{\nu}(T_{\rm ex})-J_\nu(T_{\rm bg})]  (1-e^{-\tau (v)}) e^{-\tau_{\rm dust}}. 
    \end{aligned}
\label{eq:rad_one}
\end{equation}
Here, in an optically thin case, $\tau_{\rm dust}$ is 0, i.e., $e^{-\tau_{\rm dust}}$ is $\sim$1, so that the equation reverts to the commonly used equation. Here $\tau (v)$ contains the following parameters: linewidth $\sigma$ and central velocity $V_{\rm LSR}$ as Gaussian profiles while the peak is determined by the column density $N_{\rm tot}$ and $T_{\rm ex}$ (equations \ref{eq:tau_v}, \ref{eq:tau}, and \ref{eq:boltzmann}). Together with the isotopologue ratio $N_{\rm tot, \ce{CH3CN}}/N_{\rm tot, \ce{CH3^{13}CN}}$, six free parameters are included: excitation temperature ($T_{\rm ex}$), column density ($N_{\rm tot, \ce{CH3CN}}$), dust optical depth ($\tau_{\rm dust}$), central velocity ($V_{\rm LSR}$), linewidth ($\Delta V$), and isotopologue ratio ($N_{\rm tot, \ce{CH3CN}}/N_{\rm tot, \ce{CH3^{13}CN}}$).

\subsection{Two layer LTE radiative transfer model}\label{app:lte_two}
The two layer model simply adds one more layer with independent $T_{\rm ex}$, $\tau (v)$, and $\tau_{\rm dust}$, so that:
\begin{equation}
    \begin{aligned}
    \Delta I_{\nu} (v) 
        = &\, I_{\rm line} (v) - I_{\rm dust}\\
        = &\, J_{\nu}(T_{\rm bg})e^{-[\tau_1 (v)+\tau_{\rm dust,1} +\tau_2 (v) + \tau_{\rm dust,2}]} \\
            &\, + J_{\nu}(T_{\rm ex,1})(1-e^{-[\tau_1 (v)+\tau_{\rm dust,1}]}) \\
            &\, + J_{\nu}(T_{\rm ex,2})(1-e^{-[\tau_2 (v)+\tau_{\rm dust,2}]}) e^{-[\tau_1 (v)+\tau_{\rm dust,1}]}\\
            &\, - [J_{\nu}(T_{\rm bg})e^{-(\tau_{\rm dust,1}+\tau_{\rm dust,2})} + J_{\nu}(T_{\rm ex,1})(1-e^{-\tau_{\rm dust,1}}) + J_{\nu}(T_{\rm ex,2})(1-e^{-\tau_{\rm dust,2}}) e^{-\tau_{\rm dust,1}}]\\
        = &\, [J_{\nu}(T_{\rm ex, 1})-J_\nu(T_{\rm ex, 2})]  (1-e^{-\tau_1 (v)}) e^{-\tau_{\rm dust, 1}} \\
        &\, +[J_{\nu}(T_{\rm ex, 2})-J_\nu(T_{\rm bg})]  (1-e^{-[\tau_1 (v)+\tau_2 (v)]}) e^{-(\tau_{\rm dust, 1}+\tau_{\rm dust, 2})}.
    \end{aligned}
\label{eq:rad_two}
\end{equation}
The layer one with $T_{\rm ex,1}$, $\tau_1 (v)$, and $\tau_{\rm dust,1}$ is thus the front layer while the layer two with $T_{\rm ex,2}$, $\tau_2 (v)$, and $\tau_{\rm dust,2}$ is the rear layer.
Each of $\tau_1 (v)$ and $\tau_2 (v)$ thus is described by excitation temperature ($T_{\rm ex}$), column density ($N_{\rm tot, \ce{CH3CN}}$), central velocity ($V_{\rm LSR}$), and linewidth ($\Delta V$) (see section \ref{app:lte}).
Again, with an isotopologue ratio ($N_{\rm tot, \ce{CH3CN}}/N_{\rm tot, \ce{CH3^{13}CN}}$) for each component, the two layer model has in total $6\times2=12$ free parameters.
Here again, if $\tau_{\rm dust,1}=0$ and $\tau_{\rm dust,2}=0$, the equation \ref{eq:rad_two} can be rewritten to the widely used two-layer model, i.e. $\Delta I_{\nu} (v) = J_{\nu}(T_{\rm ex,1}) [1-e^{-\tau_1(v)}] +J_{\nu}(T_{\rm ex,2}) [1-e^{-\tau_2(v)}]e^{-\tau_1(v)} - J_{\nu}(T_{\rm bg})[1-e^{-\tau_1(v)-\tau_2(v)}]$ for molecular cores (i.e., Equation 2 in \citealt{my96}). On the other hand, if the first layer (front-layer) is optically thin ($\tau_{\rm dust,2}\sim0$) and the second layer (rear-layer) is optically thick ($\tau_{\rm dust,2}\gg1$), the equation is similar to that with gas-dust decoupled model $\Delta I_{\nu} (v) = [J_{\nu}(T_{\rm ex, 1})-J_\nu(T_{\rm ex, 2})]  (1-e^{-\tau_1 (v)})$ with $T_{\rm ex,2}$ representing the dust temperature of the optically thick continuum background.

\section{Degeneracy and model selection} \label{app:degeneracy}
Given the 12 free parameters in the two-layer model, degeneracy can occur during fitting. We focus on the determination of the front and rear layers. Taking an extreme example as position 3 in Figure \ref{fig:spe_multi}, with the clearly seen double peak, the front and rear layers only have small overlap in velocity. In such a case, the fit with front layer in red-shifted and rear layer in the blue-shifted indicates an infall motion. However, the fitting can also converge to a solution with the opposite result as outward motion depending on the initial guess for the fitting or the prior set in MCMC sampling. As a result, the relative velocity can be obtained but it is difficult to determine an infalling or outward motion. We note that for the classical two-layer model in \citet{my96} or its families (see also appendix \ref{app:lte_two}), the infall signature is obtained because it assumes symmetric structures, i.e., the two layers share the same $\tau$ and linewidth. In our case, the two layers are clearly not symmetric, for example toward position 6 (Figure \ref{fig:spe_multi}), the linewidths are quite different (0.9 and 1.5 km s$^{-1}$) in the two layers. In this work, we set the initial guess to fit the spectra with blue excess using infall and with red excess using outward motion as the classical picture. 

\begin{figure*}
\includegraphics[width=0.99\textwidth]{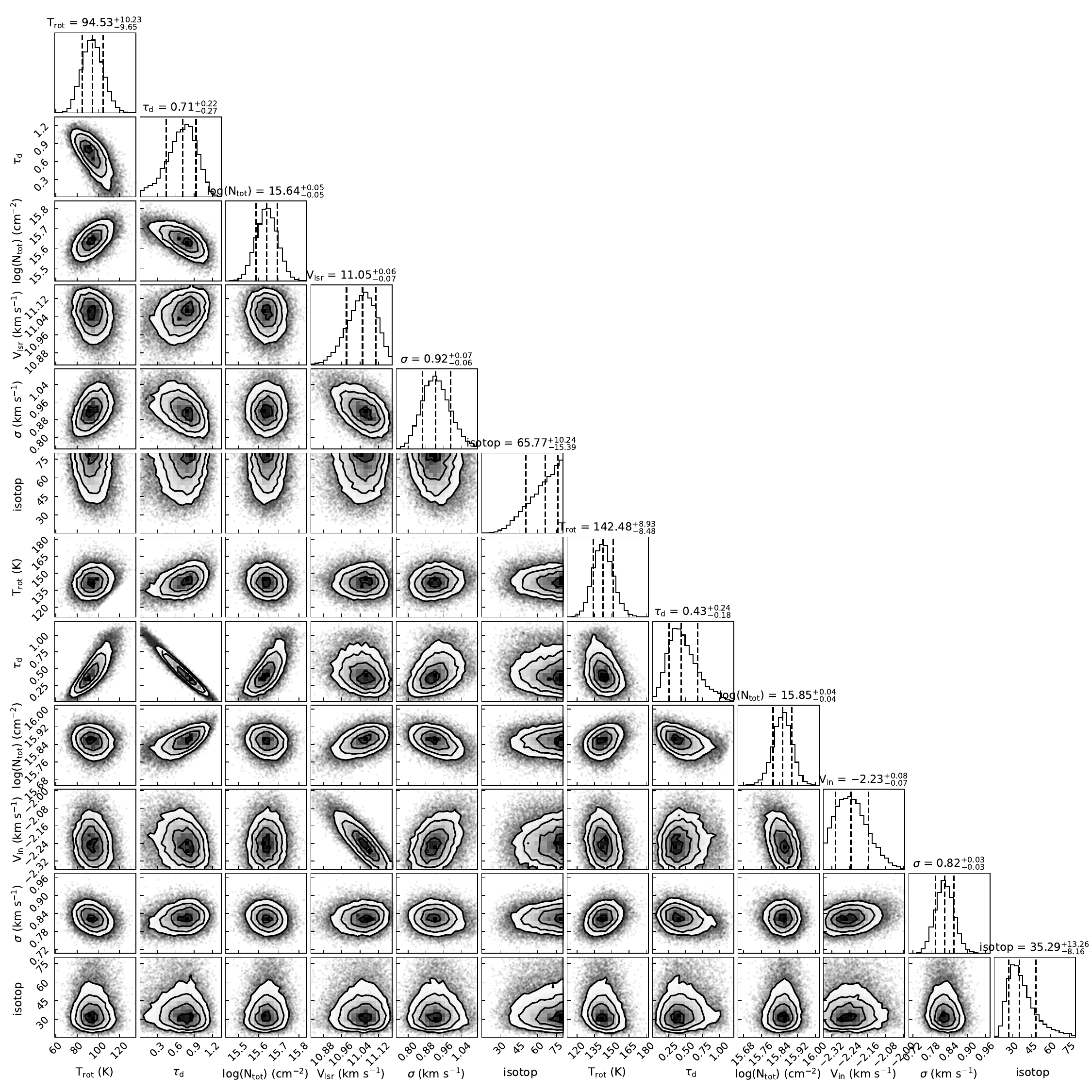}
\caption{Corner plot for the MCMC sampling toward position 1 in Figure \ref{fig:spe_multi} as a typical example. We note the 10th column represent the relative velocity between the layer 1 (front) and the layer 2 (rear).}
\label{fig:corner}
\end{figure*}

\begin{figure*}
\includegraphics[width=0.99\textwidth]{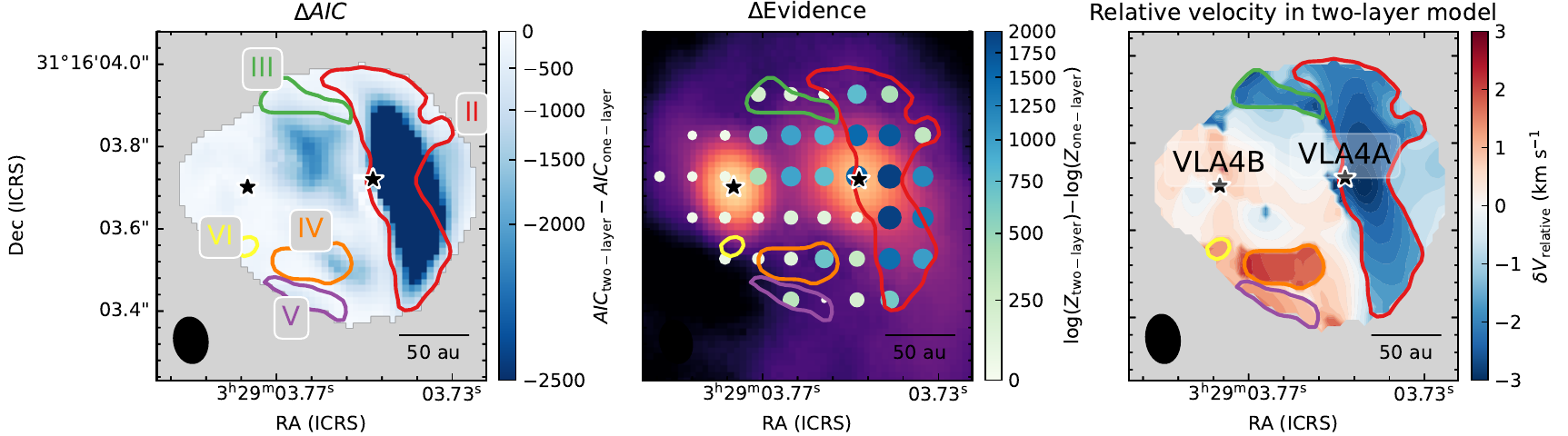}
\caption{(Left) The $\Delta AIC$ map between the one-layer and two-layer models. The color contours show the area where the two-layer model is selected and the colors indicate the kinematic components II-VI (I is the main component covering the whole map, see Figure \ref{fig:3d_cluster}). (Middle) Difference of evidences between the two-layer and the one-layer models toward the points with nested sampling overlaid on continuum image. The colors and sizes represent the difference. (Right) The relative velocity, i.e., the velocity difference $V_{\rm LSR, 2}-V_{\rm LSR, 1}$ of the layer one (front layer, $\tau_{1}(v)$) and layer two (rear layer, $\tau_{2}(v)$) in equation \ref{eq:rad_two}, in the two-layer model.}
\label{fig:aic}
\end{figure*}

Figure \ref{fig:corner} shows the corner plot from the MCMC sampling toward position 1 in Figure \ref{fig:spe_multi}. A correlation between the $\tau_{\rm dust, 1}$ and $\tau_{\rm dust, 2}$ from the two layers is shown. This might be expected given the equation \ref{eq:2_layer} or \ref{eq:rad_two}. Figure \ref{fig:para_multi} second column with $\tau_{\rm dust}$ maps also gives a hint for this degeneracy. Likely, the model shows degeneracy between $\tau_{\rm dust, 1}$ and $\tau_{\rm dust, 2}$ while the summation of these two values might be reasonable.
To overcome this problem requires observations at longer wavelengths of \ce{CH3CN} with optically thin dust continuum. Also, for the region with optically thin line emission, i.e., $(1-e^{-\tau(v)})\approx{\tau(v)}$, the constraint for $\tau_{\rm dust}$ can be more difficult given equation \ref{eq:rad_one}.
To break these degeneracies, multiple wavelength observations from optically thin to thick continuum with \ce{CH3CN} might help with understanding the dust properties. This could eventually distinguish between infall and outward motions in the spectra.

Figure \ref{fig:corner} reveals another unconstrained parameter in the isotopologue ratio of \ce{CH3CN}/\ce{CH3^{13}CN} for which the sample reaches its upper boundary of 80 in the prior. This is also expected toward the region with low S/N in \ce{CH3^{13}CN} (this example is position 1 in Figure \ref{fig:spe_multi}). Thanks to the high-sensitivity data, it is not a problem in the regions with robust \ce{CH3^{13}CN} detection (Figure \ref{fig:13_mom0}).

Here we discuss our selection of model for each pixel with either one- or two-layer. Figure \ref{fig:aic} (left panel) shows the Akaike information criterion (AIC) map ($AIC = 2{\rm k}+\chi^2+C$ where k is the degree of freedom, $\chi^2$ is the chi-squared, and C is a constant). The low $\Delta AIC$ would be in favor for two-layer model. In general, two-layer model works better, especially for the component II region. 
In addition, we have also calculated Bayesian evidence with nested sampling algorithm \citep{sk04} using Dynesty \citep{sp20,ko22} toward the 41 grid positions (Figure \ref{fig:aic} middle panel); Bayesian evidence has been used to make model selection in line modeling \citep{so20}.
The result is roughly consistent with AIC method favor two-layer model especially for the component II region.
However, toward some regions, the relative velocity of the two layers can be very small. Kinematically, this region will not be decomposed to an additional component when clustering in the position-position velocity space (section \ref{sec:clustering}).
Thus, for these regions without a clear double-peak revealed, we decide to be conservative and use the one-layer model. As a result, we use the criteria that $|\delta V_{\rm relative}|$ is larger than the velocity dispersion from both layers.

\section{Position-Velocity diagram}
\begin{figure*}
\begin{center}
\includegraphics[width=0.97\textwidth]{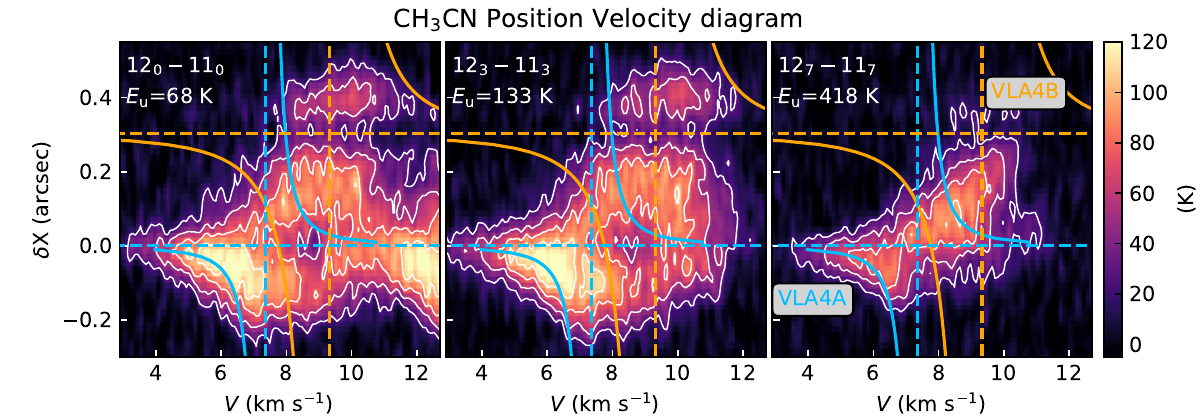}
\end{center}
\caption{PV diagram of three selected transitions from \ce{CH3CN} across VLA4A and VLA4B (west to east). The contour levels are at 5, 10, 15, and 20$\sigma$. The dashed lines indicate the position and velocity of VLA4A (blue) and VLA4B (orange). The solid curves show Keplerian profiles for each with a M$_{\rm star}=0.27M_\odot$ and $\theta_{\rm inc}=22^\circ$ for VLA4A (blue) and M$_{\rm star}=0.60M_\odot$ and $\theta_{\rm inc}=40^\circ$ for VLA4B (orange) according to \citet{di22}.}
\label{fig:12_pv}
\end{figure*}

\begin{figure*}
\begin{center}
\includegraphics[width=0.97\textwidth]{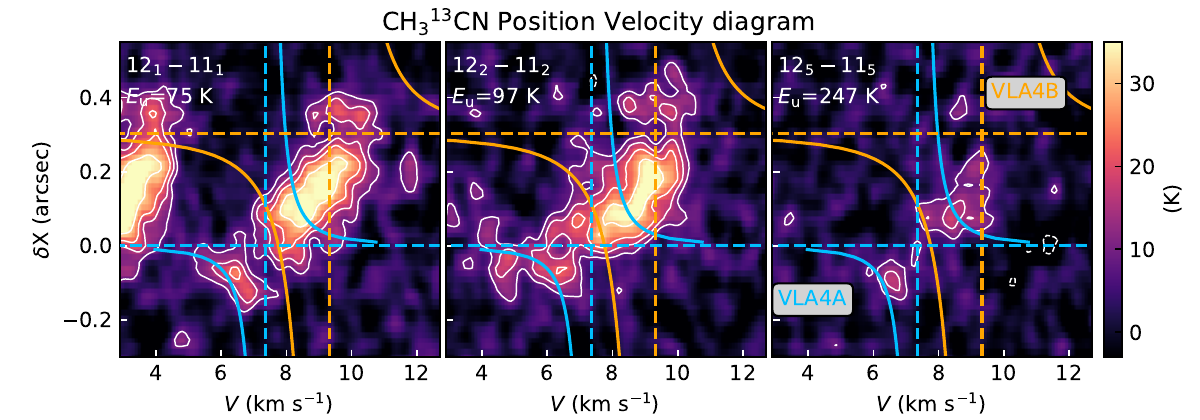}
\end{center}
\caption{Same from Figure \ref{fig:12_pv} but for \ce{CH3^{13}CN} with a contour levels of 3, 5, 7, and 10$\sigma$. To improve the S/N for the \ce{CH3^{13}CN}, the data is smoothed along the spectral axis.
}
\label{fig:13_pv}
\end{figure*}

Figures \ref{fig:12_pv} and \ref{fig:13_pv} show the position-velocity diagram of three selected transitions from \ce{CH3CN} and \ce{CH3^{13}CN}, respectively. The PV-cut is taken along the west to east along VLA4A and VLA4B. The Keplerian rotation curves are plotted given the parameters from \citet{di22} (see also Figure \ref{fig:Q}). We note that since the gas disk is not found in \citet{di22}, the inclination of the Keplerian curve is taken from the dust disk.

\end{document}